\documentclass[useAMS,usenatbib,a4]{mn2e}
\usepackage{psfig} 
\usepackage[dvips]{graphicx}
\usepackage{natbib}

\def\rxte{{\sl RXTE}}

\def\mkn79{{Mrk~79}}

\def\mch{M$\rm^{c}$Hardy\,}

\def\ecs{ergs cm$^{-2}$ s$^{-1}$~}
\def\lesssim{\mathrel{\hbox{\rlap{\hbox{\lower4pt\hbox{$\sim$}}}\hbox{$<$}}}}
\def\gtrsim{\mathrel{\hbox{\rlap{\hbox{\lower4pt\hbox{$\sim$}}}\hbox{$>$}}}}

\newcommand{\be}{\begin{equation}}
\newcommand{\ee}{\end{equation}}

\title[Optical/X-ray variability of \mkn79]{Long term Optical and X-ray
Variability of the Seyfert Galaxy Markarian 79}  
\author[E. Breedt et al.]{E. Breedt$^{1}$\thanks{E-mail:
ebreedt@astro.soton.ac.uk}, P. Ar\'evalo$^{1}$, I. M. \mch$^{1}$, P.
Uttley$^{1}$, S. G. Sergeev$^{2,3}$, \and T. Minezaki$^{4}$, Y.
Yoshii$^{4,5}$, C. M. Gaskell$^{6}$, E. M. Cackett$^{7}$, K. Horne$^{8}$, S.
Koshida$^{9}$\\ 
$^1$School of Physics and Astronomy, University of Southampton, Southampton SO17 1BJ, UK\\
$^2$Crimean Astrophysical Observatory, P/O Nauchny, Crimea 98409, Ukraine\\
$^3$Isaac Newton Institute of Chile, Crimean Branch, Ukraine\\
$^4$Institute of Astronomy, School of Science, University of Tokyo, 2-21-1
Osawa, Mitaka, Tokyo 181-0015, Japan\\
$^5$Research Centre for the Early Universe, School of Science, University of
Tokyo, 7-3-1 Hongo, Bunkyo-ku, Tokyo 113-0033, Japan\\
$^6$Department of Astronomy, University of Texas, Austin, TX 78712-0259, USA\\
$^7$Chandra Fellow, Department of Astronomy, University of Michigan, 500 Church Street, Ann
Arbor, MI 48109, USA\\
$^8$School of Physics and Astronomy, University of St. Andrews, KY16 9SS, Scotland, UK\\
$^9$Department of Astronomy, School of Science, University of
Tokyo, 7-3-1 Hongo, Bunkyo-ku, Tokyo 113-0033, Japan\\
}

\begin{document}
\date{Received /Accepted}
\pagerange{\pageref{firstpage}--\pageref{lastpage}} \pubyear{2008}

\maketitle
\label{firstpage}
 
\begin{abstract}
We present the results of concurrent X-ray and optical monitoring of the Seyfert~1 galaxy \mkn79 over a period of more than five years. We find that on short to medium time-scales (days to a few tens of days) the 2--10 keV X-ray and optical $u$ and $V$ band fluxes are significantly correlated, with a delay between the bands consistent with zero days. We show that most of these variations may be well reproduced by a model where the short-term optical variations originate from reprocessing of X-rays by an optically thick accretion disc. The optical light curves, however, also display long time-scale variations
over thousands of days, which are not present in the X-ray light curve. These optical variations must originate from an independent variability mechanism and we show that they can be produced by variations in the (geometrically) thin disc accretion rate as well as by varying reprocessed fractions through changes in the location of the X-ray corona.
\end{abstract}


\section{Introduction} \label{sec:intro}

Active Galactic Nuclei (AGN) emit across almost the entire
electromagnetic spectrum. In most AGN, the bulk of the emission is
radiated in the UV/optical region in what is known as the `big
blue bump'. The blue bump is generally accepted to be thermal
emission, originating in an optically thick accretion disc surrounding
the central black hole. For a review, see \citet{koratkarblaes99}. AGN have been known to exhibit optical variability since shortly after their discovery
\citep[e.g.][]{smithhoffleit63} but the origin of the variability
is still unclear. The two main possibilities are that
either the optical variations arise from intrinsic variations of the
disc emission, perhaps due to accretion rate variations, or that the
variations arise from reprocessing of the variable X-ray emission, boosting the intrinsic thermal emission
from the disc.

If the optical variations arise from reprocessing of X-rays, we expect them to lag behind the X-ray variations. As the X-rays originate from
close to the central black hole and the longer wavelength emission comes from
further out in the disc, light travel time arguments for a standard optically
thick disc predict that the lag will increase with wavelength as $\tau\propto\lambda^{4/3}$ \citep{collieretal99}. For a typical AGN we 
expect short delays, of order a day or less, between adjacent photometric bands.
Such wavelength-dependent delays have been seen in many AGN
\citep{wandersetal97, collier98, krissetal00, collieretal01, oknyanskij03, sergeevetal05, doroshenkoetal06}. For a detailed discussion of the disc reprocessing model, see \citet{cackettetal07}. \citet{gaskell07} suggested that the wavelength-dependent delays in the UV-optical may also be due to a contribution from reprocessing material much further out in the torus.

On the other hand, if the optical variations arise from variations in the intrinsic disc emission as accretion rate perturbations propagate inwards, we expect the optical variations to precede the X-ray variations. By the same argument, the longer wavelength optical emission will also lead those at shorter wavelengths.  The variations will propagate through the disc on a viscous, rather than light-travel time-scale, so we expect much longer delays ($\sim$months--years) between wavebands than in the reprocessing scenario.

Coordinated X-ray and optical monitoring observations provide a method
for determining the sign of the delay between the optical and X-ray
variations and hence for determining the dominant process underlying
optical variability. However, previous studies have yielded varied and
confusing results, with some sources showing well correlated emission
and others no correlation at all. For example, \citet{mason02} report
a 0.14~day lag of the UV emission relative
to the X-rays in NGC~4051, but on longer time-scales \citet{shemmer03}
find a 2.4 day optical lead. \citet{peterson00} find a correlation
between the X-rays and optical on time-scales of a few weeks in the
same source, but little agreement on time-scales of days.  For
NGC~3516, \citet{maoz00} reported a possible 100~day X-ray to optical
lag after their 1.5~year monitoring of this source. However, with the
addition of 3.5~years more data \citep{maoz02}, they were unable to
confirm this lag, or measure any other statistically significant
correlation at any lag in the light curves. For 3C~390.3, the relationship between passbands is also complicated: X-ray flares sometime lead and sometimes follow optical/UV events \citep{gaskell08}. \citet{uttley03} found
the X-ray and optical continuum in NGC~5548 to be well correlated on
month time-scales, and also that the amplitude of the optical
variations was larger than that of the X-rays on these long time-scales. On
short time-scales \citet{suganumaetal06} report a clear 1.6~day optical lag to
the X-ray variations and an optical amplitude much smaller than the X-ray
amplitude, suggesting that, unlike on the longer time-scales, reprocessing is
the main driver of the optical variability  on time-scales of days. Intensive monitoring by \citet{gaskell06}, however, fails to show this short time-scale correlation.

Besides simply measuring the optical--X-ray lag, coordinated X-ray and
optical monitoring observations enable us to put other useful
constraints on the X-ray and optical emission processes and on the
geometry of the emitting regions. For example, the greater fractional variability
of the X-rays on short ($\sim$day) time-scales indicates that, in a
Compton scattering scenario for the X-ray production, at least part of
the X-ray variability is not produced by variable up-scattered seed
photons, but by variations in the scattering corona itself. The relative
amplitude of optical and X-ray variability is also important because
if the optical variations, which represent variations in the
bolometrically dominant blue bump, exceed those in the X-rays, it is
hard to explain the optical variations as purely due to
reprocessing, e.g. as in NGC~5548 \citep{uttley03} and MR2251-178
\citep{arevalo08}. In these AGN the large amplitude optical
variations occur on time-scales of years and may represent accretion
rate fluctuations propagating through the emission region.

It has been suggested \cite[e.g.][]{uttley03} that the strength of the
correlation may be dependent on the mass of the central black hole. Scaled in
terms of the gravitational radius, AGN of higher black hole mass or lower
accretion rate, with cooler discs, will have the thermal optical emitting region
closer to the black hole than the less massive/high accretion rate objects. We
may expect stronger correlations in the high mass systems, as the accretion rate
variations will be better connected between the X-ray and optical emitting
regions. Conversely, in the low mass/high accretion rate systems, the very
different local viscous time-scales will allow the emission regions to vary
independently, causing poorer correlations on long time-scales. 

In order to try and understand the optical variability of AGN, and the
relative importance of the two proposed variability mechanisms, we
have set up a long term monitoring programme of a sample
of AGN of differing mass and fractional accretion rate $\dot{m}=\dot{M}/\dot{M}_{\mbox{\small{Edd}}}$. Previously \citep{uttley03} we have
reported on observations of NGC~5548 
[$M\sim 7\times10^{7}M_\odot$; $\dot{m}\sim0.08$ \citep{peterson04,woourry02}] 
and MR2251-178 [$M\sim10^{9}M_\odot$;  $\dot{m}\sim 0.04$] \citep{arevalo08} and here we report on the observations of \mkn79.

\begin{figure} 
\psfig{file=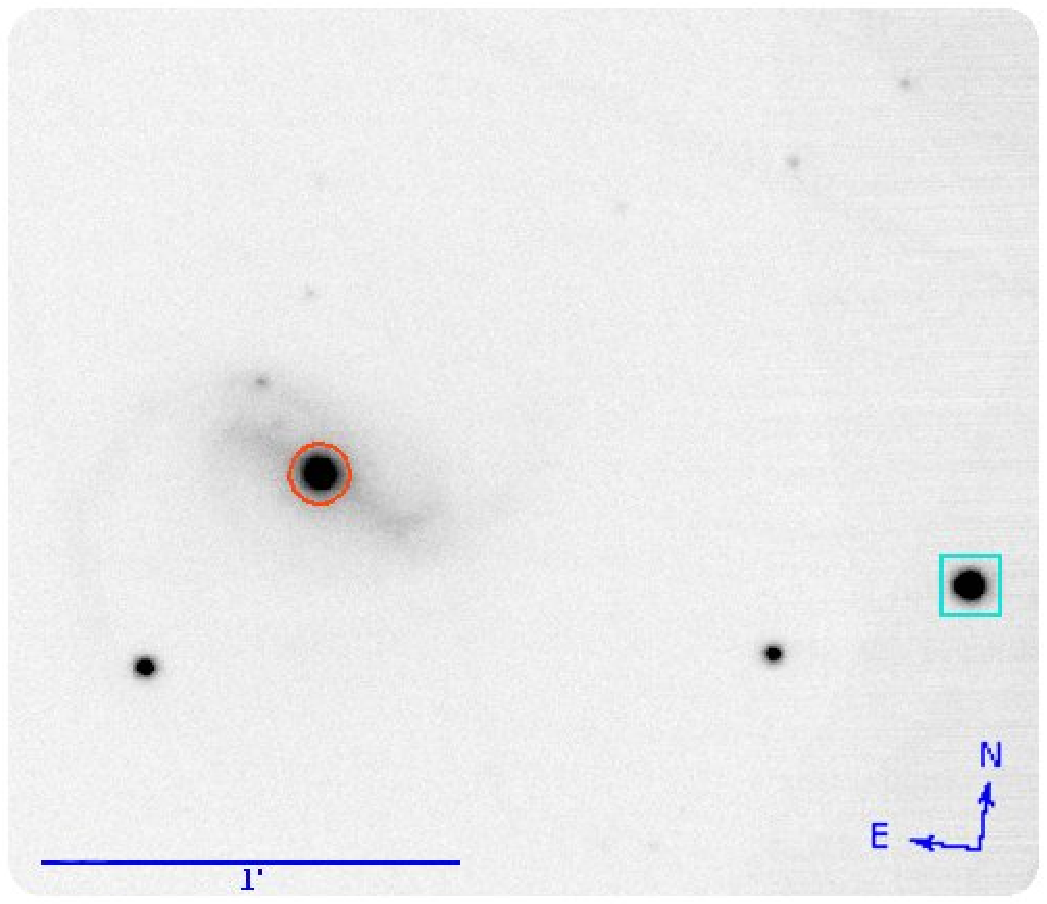,width=8cm}
    \caption{A combined $V$ band image of the Seyfert galaxy Markarian 79 (marked with a red circle) as obtained with the Liverpool Telescope. The bar at the bottom of the image respesents a scale of 1 arcminute. The comparison star used for the optical photometry is marked with a blue square, and the summed flux from the other two bright stars in the image was used to confirm that the comparison star did not vary.}
    \label{fig:pic}
\end{figure}

\mkn79 (UGC~3973, MCG+08--14--033) is a nearby ($z=0.022$) Seyfert~1.2
galaxy with a black hole mass of $(5.24\pm1.44)\times 10^7 M_\odot$
\citep{peterson04} and an average $\dot{M}$ of $\sim6\%$ of its Eddington accretion rate \citep{kaspi00}. A $V$ band image of the galaxy is shown in Figure~\ref{fig:pic}. It is a well-known X-ray source and its optical variability is well established
\citep{peterson98,webbmalkan00}. We present the combined optical light
curves from four independent monitoring programs, together spanning
more than five years, as well as the X-ray monitoring light curve as
obtained by the \rxte~satellite for the same time. Data collection,
construction of the light curves and basic variability properties are
discussed in Section~2. We present the cross-correlation analysis in
Section~3 and in Section~4 we show that the medium term fluctuations
are well reproduced by a reprocessing model and that the long
time-scale trends can be accounted for by allowing the disc and/or
corona geometry to vary over these time-scales. We conclude in
Section~5.

\section{Observations and data reduction} \label{sec:data}

\subsection{X-ray observations}
We used the Proportional Counter Array (PCA) on board \rxte~to obtain a long term X-ray light curve of \mkn79. From 2003 March 23 to 2008 January 30, we took snapshots of $\sim1$~ks exposure time each, at two day intervals. During a period of intensive monitoring, 2005 November 17 to 2006 January 19, four observations per day were made. Only data from PCU$\:$2 were used, as PCUs 1, 3, and 4 were regularly switched off during our observations and PCU$\:$0 has known problems with high background since the loss of its Xenon layer. The data were reduced and analysed with {\sc FTOOLS} v.6.4, using standard extraction methods and acceptance criteria. The background was calculated from the most recent background models which corrects for the recent problems with the \rxte~SAA history file. The final 2--10 keV fluxes were calculated by fitting a power law to the observed spectra. This allows one to take account of changes in the instrumental gain over the duration of the monitoring.

\begin{figure*} 
\rotatebox{270}{\includegraphics[width=8cm,height=18cm]{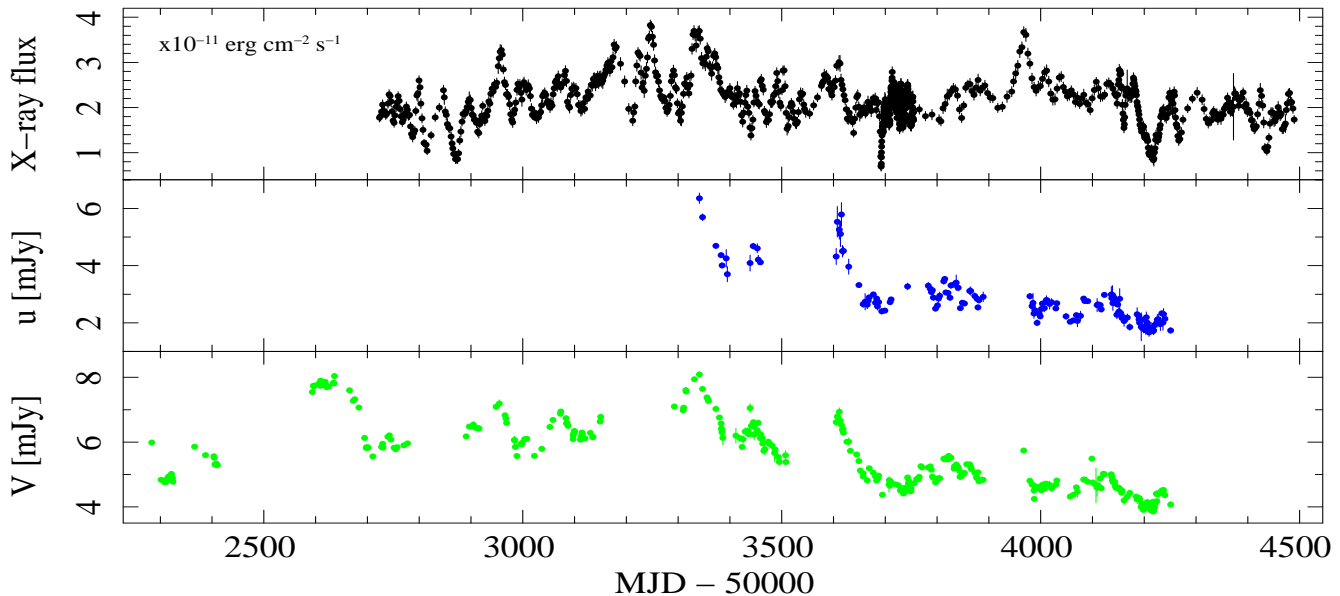}}
    \caption{$2-10$ keV X-ray (top panel), $u$ (middle panel) and $V$ band (bottom panel) light curves of \mkn79. The X-ray light curve is smoothed with a 4 point ($\sim$8~day) boxcar, to highlight the variations on time-scales similar to that of the optical light curves. The X-ray light curve is in units of $10^{-11}$ \ecs, the optical light curves are in mJy. The $u$ band flux is scaled to a comparison star in the field but no colour correction has been applied. $V$ flux calculated as described in the text. No host galaxy contribution has been subtracted.\label{fig:xuv}}
\end{figure*}

The X-ray light curve shown in Figure~\ref{fig:xuv} is smoothed with a 4~point (approximately 8~day) running average to show the variations on time-scales of days more clearly. However, the unsmoothed light curve was used in all calculations.

\subsection{Optical observations}
Since 2002, \mkn79 has been monitored independently by several ground-based observatories. We combine the data from six different telescopes here to construct an optical light curve spanning more than 5 years. A summary of these observations may be found in Table~\ref{tab:optsummary} and each is discussed briefly below.

\begin{table*}         
   \caption{Summary of $V$ band observations \label{tab:optsummary}}
   \begin{tabular}{c c c c}
      \hline
      Observation dates & & Aperture & No. of $V$ band \\
       (MJD) & Instrument and Telescope & diameter & observations \\
      \hline
      52283.5 -- 53987.6 & AP7p camera, 0.7-m Crimean Observatory Telescope & $15''$ & 142\\
      52890.6 -- 53876.3 & Multicolor Imaging Photometer, MAGNUM Telescope & $8\arcsec.3$ & 47\\
      53100.6 -- 54198.8 & 0.4-m telescope, Lincoln, Nebraska & $16''$ & 79\\
      53341.0 -- 54250.9 & RATCam, Liverpool Telescope & $12''$ & 132\\
      54063.4 -- 54129.3 & 0.6-m telescope, Mt. Maidanak Observatory & $16''$ & 22\\
      54136.3 -- 54237.3 & HawkCam, Faulkes Telescope North & $12''$ & 39\\
      \hline
   \end{tabular}
\end{table*}

\subsubsection{Liverpool Telescope}
\mkn79 is part of our ongoing optical monitoring programme on the Liverpool Telescope (LT), a 2 metre fully robotic telescope at the Observatorio del Roque de Los Muchachos of the Instituto de Astrofisica de Canarias, La Palma, Spain \citep{livtel}. The telescope is equipped with an optical CCD camera, RATCam, with a 2048$\times$2048 pixel back-illuminated detector and a scale of 0.135 arcsec/pixel. Observations in the SDSS-$u$ and Bessell-$V$ filter bands started on 2004 December~2 with a four day sampling period. This was increased in March 2007 to daily monitoring in $V$ and every two days in $u$. The latest measurement in the light curve included here was taken on 2007 May 30.

The data are overscan-trimmed and flat-fielded by the LT reduction pipeline. 
Bias correction is based on the underscan region only, as the bias frames do not show any significant repeatable structure. Images are binned 2$\times$2, resulting in a 1024$\times$1024 pixel image of the 4.6~arcminute field of view. Standard rejection criteria are applied: data from nights with poor seeing ($>3\arcsec$) or a very bright moon were discarded. The median seeing over the period of monitoring was $1.3\arcsec$. A few hot pixels and one bad column on the CCD were removed by linearly interpolating between adjacent pixels.

\begin{figure} 
    \rotatebox{270}{\includegraphics[width=6cm,height=8cm]{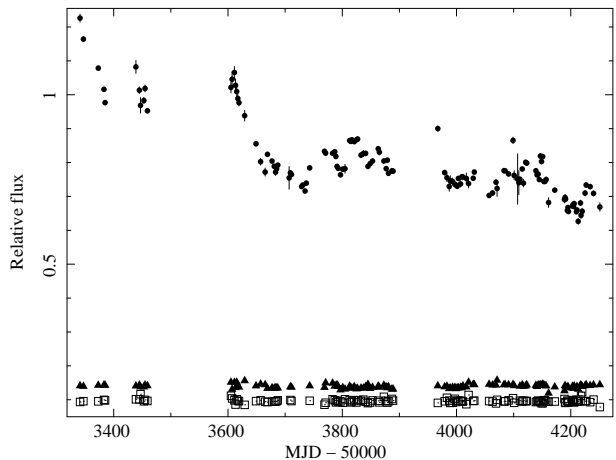}}
    \caption{$V$ band light curve of \mkn79 as obtained with the Liverpool Telescope. The ratio of the comparison star to two check stars in the field is also shown, showing that the decrease in brightness of \mkn79 is real and not due to a variable comparison star.}
    \label{fig:LT_V}
\end{figure}

We used the IRAF\footnote{IRAF is distributed by the National Optical Astronomy Observatories, which are operated by the Association of Universities for Research in Astronomy, Inc., under cooperative agreement with the National Science Foundation} package DAOphot to perform aperture photometry on the images. We summed the flux inside a $12\arcsec$ diameter aperture, centred on the nucleus, and compared it to a comparison star and 2 check stars in the field. Comparison star fluxes were taken from \citet{doroshenko05i} and \citet{mihovslavcheva07}. The aperture size was selected to include the PSF under seeing conditions of $3\arcsec$ or less.  In order to exclude host galaxy flux from the background measurement, the sky background level was not measured in an annulus. Instead, we measured and averaged the background from three $12\arcsec$ diameter apertures, displaced in various directions around \mkn79 and away other stars in the field. Aligning all of our images to a reference image before these measurements were taken, ensured that the same sky areas were measured in each image. Photometric errors are based on photon statistics and the readout noise of the detector. \mkn79 showed a steady decline in brightness over the period of monitoring. The $V$ band light curve from the LT is shown in Figure~\ref{fig:LT_V}, relative to the comparison stars. 

\subsubsection{Faulkes Telescope}
\begin{figure} 
\begin{center}    \rotatebox{270}{\includegraphics[width=11cm,height=8.5cm]{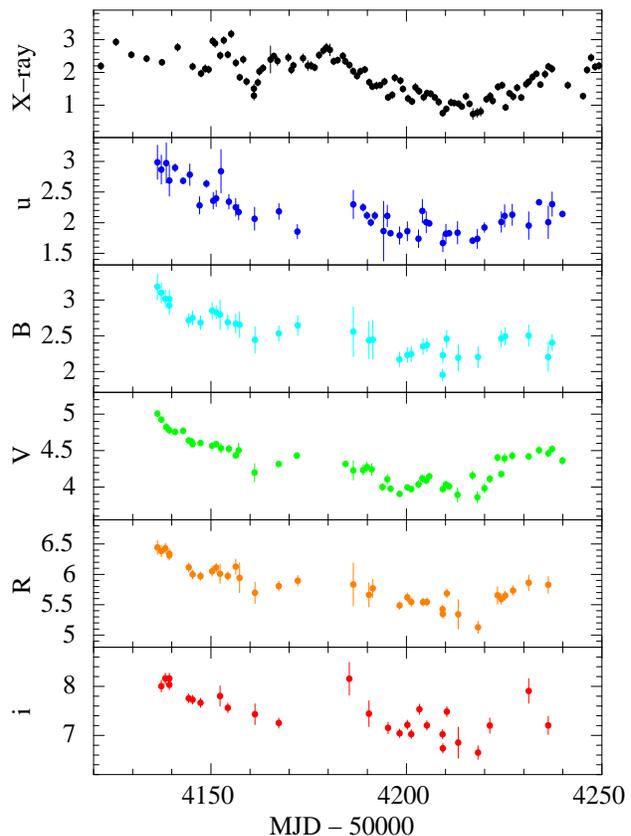}}
    \caption{$uBVRi$ light curves of \mkn79 as obtained with the Faulkes Telescope, shown with the corresponding section of the X-ray light curve. The X-ray light curve is in units of $10^{-11}$ \ecs, the optical light curves are in mJy.}
    \label{fig:xubvri}
\end{center}
\end{figure}

We also carried out a short multicolour monitoring program in $u$,$B$,$V$,$R$ and $i$ on the Faulkes Telescope North (Haleakala, Maui, Hawaii) from 2007 February 5 to May 17. The telescope and camera are identical in specification and design to the Liverpool Telescope and data were reduced and analysed in the same way. The sampling is roughly every two days, with some gaps due to poor weather conditions. 33 useful epochs in $u$, 35 in $B$, 39 in $V$, 36 in $R$ and 27 in $i$ were obtained. The light curves are shown in Figure~\ref{fig:xubvri}.

\subsubsection{Crimean Observatory Telescope}
We supplement these light curves with $V$ band data from the AGN monitoring program on the 0.7-m telescope of the Crimean Astrophysical Observatory, described in \citet{sergeevetal05}. The $B,V,R,R1,I$ filter set on the AP7p camera mounted on this telescope are nonstandard filters, though the $V$ filter, from which we  include data here, matches the standard Bessell-$V$ filter closely. The light curve is constructed from aperture photometry through a $15\arcsec$ diameter aperture centred on the galaxy nucleus, relative to a comparison star in the field. It covers the period 2002 January~9 -- 2006 September~9 and has a sampling rate of roughly every two days. Data after MJD 53058 (2004 January 18) from this program was previously unpublished.

\subsubsection{MAGNUM Telescope}
We also include unpublished data from the MAGNUM (Multicolor Active Galactic Nuclei Monitoring) project \citep{yoshii03magnum}. The 2~metre telescope is situated at the University of Hawaii's Haleakala Observatory, on the island of Maui, Hawaii, and monitors a large sample of AGN in optical and infrared wavebands. The $V$ band light curve included here are the observations made during the period 2003 September 8 -- 2006 May 2. The nuclear flux was measured within a $8\arcsec.3$ diameter aperture, relative to a comparison star in the field. Comparison stars were calibrated against photometric standard stars and colour corrections applied to the fluxes. Full details of the data aquisition and reduction are described in \citet{suganumaetal06}.

\subsubsection{0.6-m telescope, Mt. Maidanak}
\mkn79 was also observed with the 0.6-m telescope at the Mt. Maidanak Observatory, Uzbekistan from 2006 November 24 to 2007 January 29. Aperture photometry was performed on these data using a $16\arcsec$ diameter aperture and a background annulus between $24-40\arcsec$. The measurements were calibrated using the \citet{doroshenko05i} comparison star magnitudes.

\subsubsection{0.4-m telescope, Lincoln, Nebraska}
Finally, we include $V$ band CCD observations obtained with the 0.4-m telescope of the University of Nebraska. Details of these observations and reduction procedures are as given in \citet{klimek04}. The photometric aperture, sky annulus, and comparison stars were as for the Mt. Maidanak observations.

\subsection{Relative calibration}
In order to account for the differences in the calibration and photometry aperture with which the individual light curves were constructed, we select the MAGNUM flux calibrated light curve as the reference dataset and scale all the other light curves to match it. We extracted pairs of measurements, taken within one day of each other, from overlapping light curves, and performed a least-squares fit of the equation \be F_1 = aF_2 + b \ee to the flux pairs. $F_1$ and $F_2$ are the flux measurements from the two datasets and $a$ and $b$ the parameters of the fit. The best-fit parameters from each pair of overlapping light curves were then used to scale that dataset to the $8\arcsec.3$ diameter aperture MAGNUM light curve. 

The total combined $V$ band light curve spans 1967 days (approximately 5~years and 5~months) and  contains 461 epochs. It is shown in Figure~\ref{fig:xuv} along with the X-ray and $u$ band light curves. Throughout the rest of this paper, we will use this combined light curve when we refer to the $V$ band light curve.

\subsection{Variability properties} \label{sec:varprop}

The flux measured within the aperture contains the nuclear flux as well as starlight from the part of the host galaxy falling inside the aperture. This host galaxy flux is a constant contribution to the total measured flux, so although it does not affect the results of the cross-correlation analyis (Section \ref{sec:ccf}), it dilutes the variability of the nucleus. 
We estimate the host galaxy contribution as follows. We obtained an archival HST image of \mkn79 from the online MAST archive, taken on 2006 November 8 (MJD 54048) with the Advanced Camera for Surveys through the F550M filter. This is a medium width continuum $V$ band filter, centred on 5580\AA~with a FWHM of approximately 540\AA. We compared the flux measured through an $8\arcsec.3$ aperture (to match the total $V$ band light curve) with the flux measured through a $0\arcsec.25$ aperture centered on the nucleus. We measure a 43.5\% nuclear contribution to the total optical emission in the $8\arcsec.3$ aperture. We linearly interpolate our $V$ band light curve to estimate the flux at the time of the HST observation and take 56.5\% of this value to be the host galaxy contribution. This equated to a 2.54~mJy correction, which was subtracted from the whole $V$ band light curve for the purposes of the variability calculations below.
No colour correction factor has been applied to transform this flux to the flux through the broader Bessell-$V$ filter. However, assuming the spectrum to be flat over the width of the filter, the correction to the flux ratio as calculated here, will be very small.

In order to compare the amount of intrinsic variability in each light curve we calculate the error-corrected flux variance, as a fraction of the mean flux. This quantity is known as the fractional variability, \be F_{\mbox{\tiny var}}= \frac{\sqrt{\sigma^2-\varepsilon^2}}{\langle f \rangle}\;,\ee %
where the flux variance has the standard definition
\be \sigma^2 = \frac{1}{N}\sum_{i=1}^{N}(f_i-\langle f \rangle)^2 \ee
and $\varepsilon^2$ is the sum of the squared measurement errors
\be \varepsilon^2 = \frac{1}{N}\sum_{i=1}^{N}\varepsilon_i^2. \ee
The sampling statistics and variability properties of each light curve is summarised in Table~\ref{tab:var}.

\begin{table*}
   \caption{Sampling and variability characteristics \label{tab:var}}
   \begin{tabular}{c c c c c c c}
      \hline
      Light curve & Total length (d) & N & $\Delta t_{\mbox{\tiny median}}$ (d) & $F_{\mbox{\tiny var}}$ & $f_{\mbox{\tiny max}}/f_{\mbox{\tiny min}}$ \\
      \hline
      X-ray       & 1775 & 975 & 1.89 & 0.2733 & 10.50 \\
      $u$ (total) &  910 & 143 & 2.47 & 0.3107 & 3.81  \\
      $V$ (total) & 1967 & 382 & 2.00 & 0.3447$^{\tiny a}$ & 4.20$^{\tiny a}$ \\
      {\em Multicolour monitoring:} & & & & \\
      $u$   &  103 &  47 & 1.75 & 0.1432 & 1.79 \\       	
      $B$   &  101 &  35 & 2.00 & 0.0968 & 1.63 \\       	
      $V$   &  103 &  51 & 1.73 & 0.0653 & 1.30 \\       	
      $R$   &  100 &  36 & 2.00 & 0.0504 & 1.26 \\       	
      $i$   &   99 &  27 & 2.93 & 0.0544 & 1.23 \\       	
      \hline
      \small{$^{a}$ Host galaxy flux subtracted.}
   \end{tabular}
\end{table*}

The X-ray and total optical light curves are both highly variable. The unsmoothed X-ray light curve contains much more variability power at high frequencies than the optical light curve, but due to the large amplitude of the long term optical variations, there is more variability power in the optical light curve when corrected for host galaxy flux.  In the much shorter Faulkes Telescope colour light curves, the fractional variability is seen to decrease with wavelength. If the variabiliy is intrinsic to the accretion disc, this behaviour is expected, as the longer wavelength emission originate from further out in the disc, where the natural time-scales are longer. This is also expected from reprocessing models: the disc response function broadens for longer wavelengths, smoothing out the variability at longer wavelenths \cite[see, e.g.][]{cackettetal07}. We note, however, that no correction for host galaxy flux has been made to these light curves, so the variability amplitude we measure is reduced by this constant offset. Thus the fractional variability quoted in Table~\ref{tab:var} should be viewed as a lower limit.

\section{Cross-correlation analysis} \label{sec:ccf}

\subsection{X-ray -- optical correlation}
\label{sec:ccfxrayopt}
In order to confirm and quantify the apparent correlation between the X-ray and
optical emission, we use the Discrete Correlation Function (DCF) method of
\citet{edelsonkrolikdcf}. 

We cross-correlate the unsmoothed, unbinned X-ray light curve with the observed
$u$ and $V$ band light curves and bin the resulting discrete correlation
function in uniform lag bins. Throughout this paper we will use a positive value on the lag axis to indicate a lag of the longer wavelength emission behind the shorter wavelength (i.e. a lag of the optical variations behind the X-rays, in this case). 

\begin{figure}
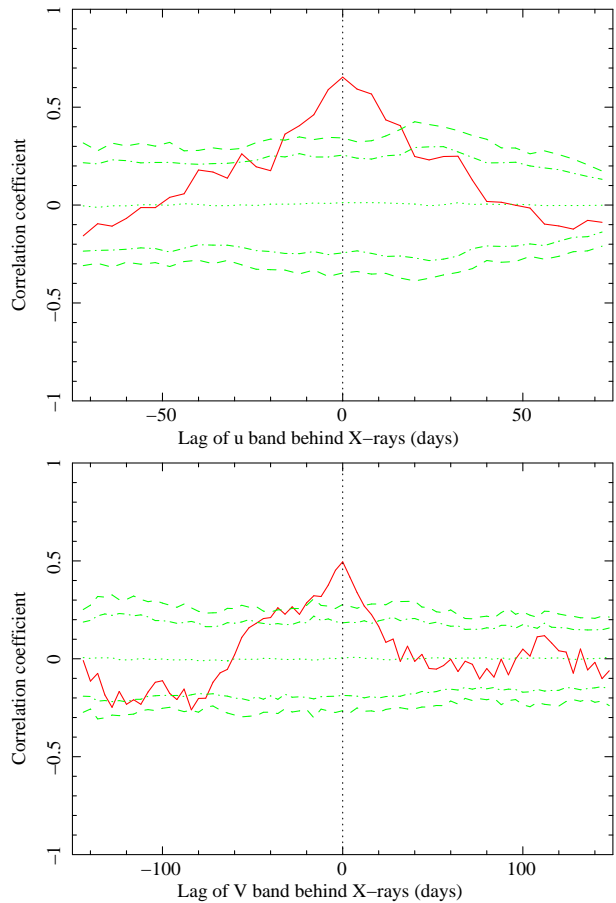

    \rotatebox{270}{\includegraphics[width=6cm,height=8cm]{dcfxu4segsims.ps}}
    \rotatebox{270}{\includegraphics[width=6cm,height=8cm]{dcfxv4segsims.ps}}
    \caption{DCFs between the X-ray and $u$ band (top) and X-ray and $V$ band (bottom), binned in 4~day lag bins. The horizontal dotted, dash-dot and dashed lines are the median, 95\% and 99\% confidence levels, based on 1000 simulations. The vertical dotted line indicates the zero lag position.}
    \label{fig:dcfxvu}
\end{figure}

To account for the $\sim$100-day long solar conjunction gaps in the optical light curves, we divide them into segments defined by the observation gaps, and calculate the DCF for each segment separately. As the stength of the correlation may be different on different time scales, we resample the X-ray light curve to a uniform observational frequency by selecting only one point every two days from the intesively sampled period starting just before MJD 53700. This ensures that this segment does not dominate the combined correlation function. The final correlation function is then the combination of the individual DCFs, weighted by the number of points in each lag bin. The DCFs calculated this way, using a bin size of 4~days, are shown in Figure~\ref{fig:dcfxvu}.

The effect of calculating the DCF in segments like this is to implicitly remove the low frequency power (i.e.\ long term trends) in the light curve, leading to an improved estimate of the short-term lag. The most common way of removing this low frequency power is to subtract a low-order polynomial function from the light curve prior to calculating the CCF \citep{welsh99}.  We prefer the method of segmentation of the light curve, however, as it also eliminates the large observation gaps from the DCF calculation. We note, however, that removing a linear trend or a higher order polynomial from the light curve also result in a correlation peak of greater than 99\% significance, as discussed below.

The correlation function peaks at DCF$_{\rm{max}}=0.496$ for the X-ray/$V$ band correlation and at DCF$_{\rm{max}}=0.654$  between the X-rays and $u$ band, with no lag between the optical bands and the X-rays. The DCF peak occurs at $0^{+4}_{-4}$ days for $V$ and $0^{+4}_{-4}$ days for the $u$ band. (Note that, because the DCF was used, these errors will always be a multiple of the bin size.) The peak of the correlation function is thought to provide the most reliable estimate of the lag \citep{welsh99}, however, when using the DCF, time resolution on time-scales shorter than the bin size is lost, so the peak position is affected by the chosen DCF binning factor.
We therefore deem the measurement of the centroid lag more reliable and stable in this case. We calculate the centroid  as the weighted average of all DCF points greater than $80\%$ of DCF$_{\rm{max}}$. For the X-ray/$V$-band correlation, the centroid lag occurs at $0.00^{+1.91}_{-4.00}$ days and for the X-ray/$u$-band at $-1.95^{+1.95}_{-3.89}$ days. The errors quoted are the 1~$\sigma$ errors calculated using the Flux Randomisation/Random Subset Selection (FR/RSS) method of \cite{peterson_frrss} and the centroids are the median centroids from 1000 such selections. 

As a separate check, we used the $z$-transformed discrete correlation function (ZDCF) of \cite{alexanderzdcf} and the interpolation cross-correlation function  \citep[ICCF;][]{gaskellsparke86,whitepeterson94} to calculate the same correlation. The correlation functions resemble those calculated using the DCF closely and the lag results are consistent between the three methods. The short lags obtained from both the peak and the centroid measurements suggest that the reprocessor must be located near the X-ray source in this object, and not as far out as the dust torus.

To test the significance of the DCF peaks found, we used the method of
\citet{timmerkonig} to simulate X-ray light curves based on an underlying model power spectrum. We use the single-bend power law parameters, measured by Summons et al. (in preparation), as appropriate for the power spectrum of \mkn79. To account for long term trends in the data, the simulated light curves were generated 10 times longer than the real light curve, and then sampled in the same way as the observed X-ray light curve. Observational noise was added at the level measured in the observed light curve. These uncorrelated, simulated X-ray light curves were then each cross-correlated with the real, observed optical light curve, to estimate the probability of obtaining equal or higher correlation peaks by chance, due to the red noise nature of the light curves. We record the mean, 95\% and 99\% extremes of 1000 simulations and plot them in dotted, dot-dashed and dashed lines, respectively, in Figure~\ref{fig:dcfxvu}. Both the $u$ and the $V$ band peaks reach higher than 99\% of the simulations, showing that the correlations are statistically significant at greater than 99\% level.

\subsection{Lags between optical bands} 
\begin{figure}
    \includegraphics[width=9cm,height=9cm]{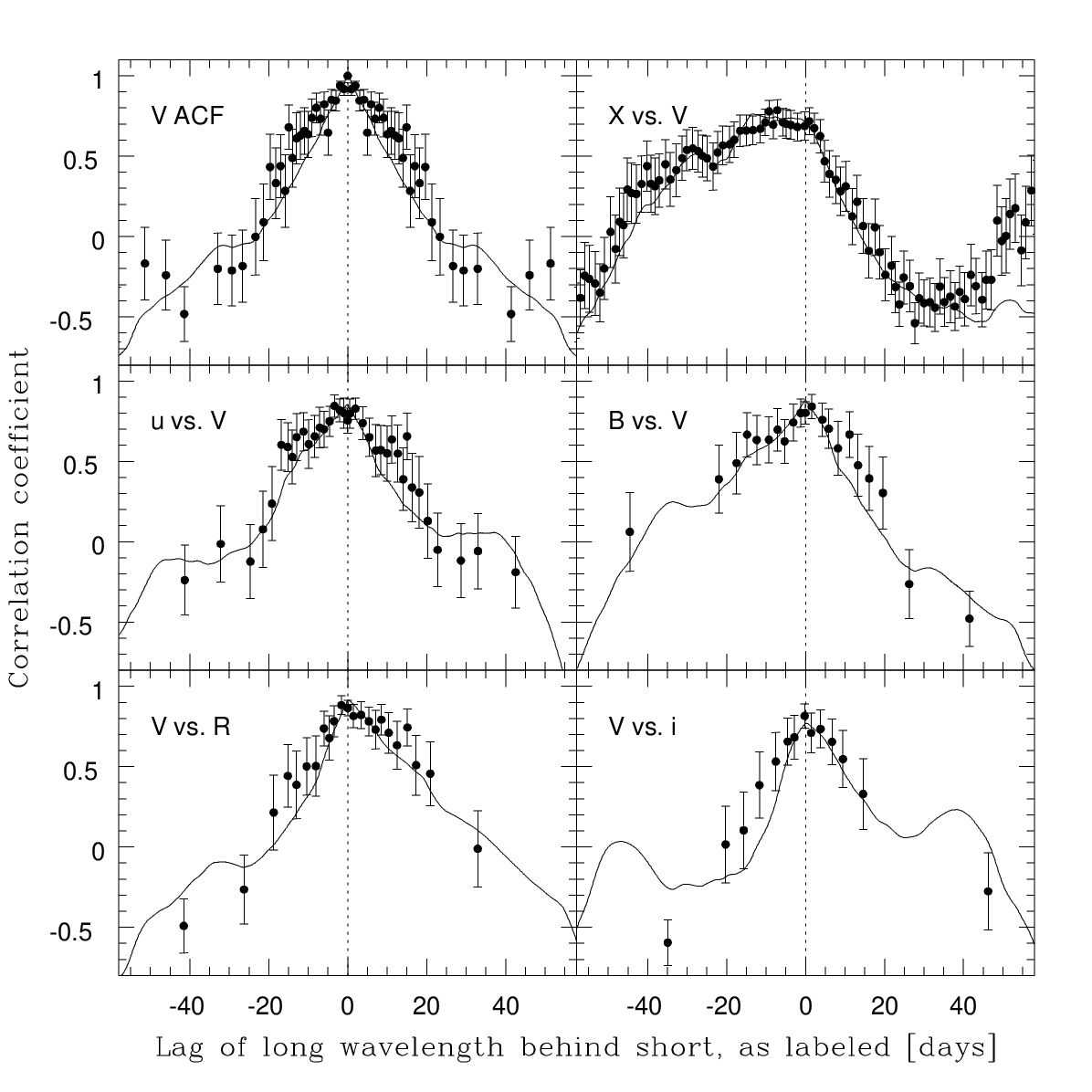}
    \caption{Cross-correlation functions relative to the $V$ band, calculated from the 3 months of Faulkes monitoring. Points with error bars are calculated from the ZDCF and the solid line from the ICCF. A positive value on the lag axis indicate a lag of the long wavelength emission behind the short wavelength. The vertical dotted lines show the position of zero lag.}
    \label{fig:zdcfvxubri}
\end{figure}

We also present, in Figure~\ref{fig:zdcfvxubri}, the ZDCF between the $X$,$u$,$B$,$R$ and $i$ bands, calculated relative to the $V$ band, for the data from the Faulkes Telescope. The $u$ and $V$ band light curves were supplemented with data from the Liverpool Telescope covering the same period.
We chose the $V$ band as reference light curve as it is the best sampled. The X-ray and $V$ band autocorrelation functions (ACF) are also shown. Again a positive value on the lag axis represent a lag of the variations in the longer wavelength band behind those of the shorter wavelength.
The interpolation cross-correlation function \citep[ICCF;][]{gaskellsparke86,whitepeterson94} is used as a consistency check. It is plotted as a solid line in Figure~\ref{fig:zdcfvxubri} and is in close agreement with the ZDCF results.

If the optical variations are due to reprocessing of X-rays by the accretion disc, standard accretion theory for a thin viscous disc allows us to calculate the lags expected between optical bands.
In a standard accretion disc model \citep[e.g.][]{fkr}, the disc can be approximated by annuli at different temperatures, each emitting a blackbody spectrum. The radial temperature profile of the disc may then be described by %
\begin{equation} T(r)=1.06\times10^6 \left( \frac{\dot{M}}{\dot{M_E}} \right)^{\frac{1}{4}} \left(\frac{M}{10^6 M_\odot}\right)^{-\frac{1}{4}}\left(\frac{r}{R_G}\right)^{-\frac{3}{4}}\mbox{ K} \label{eq:discT} \end{equation} %
where $R_G=GM/c^2$ is the gravitational radius of the black hole.
Using Planck's Law to calculate the peak emission, we can use (\ref{eq:discT}) to calculate the radius most of this emission is coming from. In the reprocessing model, delays should exist between different wavebands, due to light travel time differences between the illuminating source (assumed to be the X-rays) and the region of the disc emitting at the given wavelength \citep{collier98}. Hence variations observed at longer wavelengths should follow those at shorter wavelengths with a delay $\tau=r/c$, where $r$ is the distance between the locations of peak emission for the relevant wavelengths. Using the reverberation mapping mass \citep{peterson04} and accretion rate \citep{woourry02} of \mkn79, and assuming the X-ray source to be a point source at height $h_X=6R_g$ above the accretion disc and on its axis of symmetry, we expect a $V$ lag of 1.68 days behind the X-rays and lags between the optical bands of $u$ vs. $V=0.74$, $B$ vs. $V=0.43$, $V$ vs. $R=0.64$ and $V$ vs. $i=0.85$ days.

Both figures \ref{fig:xubvri} and \ref{fig:zdcfvxubri} show that the variations in the optical bands are well correlated. The centroids of the correlation function, calculated at 80\% of the peak, were found to be $u$ vs. $V=-0.59_{-1.92}^{+2.16}$, $B$ vs. $V=-0.58_{-2.29}^{+2.03}$, $V$ vs. $R=2.50^{+1.25}_{-1.61}$ and $V$ vs. $i=1.23^{+1.68}_{-1.49}$ days. All except the $R$ band are consistent with the expected lags for reprocessing to within the 1-$\sigma$ FR/RSS errors. The discrepancy in the $R$ band result is likely due to the strong, broad H$\alpha$ emission line contained within the $R$ filter passband. The H$\alpha$ line emission originates in the broad line region, which is further from the X-ray source than the region of the disc emitting primarily at this wavelength. Hence H$\alpha$ varies with a longer lag than the disc $R$ band emission and increases the average lag measured in this band.

From this short segment of the data, we measure the $V$ band emission to lead the X-rays by $\sim$5~days (i.e. $X$ vs. $V=-5.25_{-7.66}^{+2.55}$ days), which is not consistent with the reprocessing picture. We point out, however, that the error bars are overlapping with the $X$ vs. $V$ correlation measurement found in section \ref{sec:ccfxrayopt}. Since the sampling of these short light curves are not significantly better than the total $V$ band light curve (see Table~\ref{tab:var}), we deem the previous result of an optical lag of $0.00^{+1.91}_{-4.00}$ days behind the X-rays more robust and reliable, as it was derived from light curves almost 20 times as long. In short sections of the light curve the long term variations (which are present in the optical light curve but not the X-rays) can influence the lag measured, especially if there are not many peaks or troughs in that segment to constrain the lag tightly. In the longer light curves, these small shifts average out and a more reliable estimate of the true lag is obtained.

\section{Modelling reprocessing}

A possible mechanism for the production of optical fluctuations is thermal reprocessing of variable X-rays shining on optically thick material, possibly the accretion disc. We constructed reprocessed optical light curves, using the observed X-ray light curve as input, to compare to the observed optical light curve. We assume that the X-rays are emitted by a compact source above the accretion disc and on its axis of symmetry while the optical flux arises thermally from an optically thick accretion disc. This geometry of the X-ray source may be viewed as a first order description of the centre of an X-ray corona above the inner part of the accretion disc. More realistic geometries, such as a spherical corona of radius $h_X$, will have a small effect on the flux calculated at the inner parts of the disc, but from further out in the disc, where most of the emission we are interested in is coming from, the corona will increasingly resemble a point source. This simplistic description is thus sufficient for our purposes here. We followed the prescription detailed by \citet{reprocessing} to account for light travel time effects on the reprocessed light curve, as well as geometrical considerations on the amount of flux received per unit area for different locations in the disc and X-ray source heights. We added the impinging X-ray flux to the locally dissipated flux expected for a thin disc to calculate the black body temperature as a function of time and radius, in order to produce flux light curves in different optical bands. 

The model parameters are the height of the X-ray source $h_X$, thin disc accretion rate $\dot m$, inclination angle to the observer $i$ and inner truncation radius of the thin disc $R_{\rm in}$. The mass of the black hole determines the size of the gravitational radius and therefore the conversion of light crossing time to days. We left the mass fixed at the reverberation-mapped value of $5.24\times 10^7 M_\odot$ \citep{peterson04}. We varied $R_{\rm in}$ and $h_X$ between 6 and 50 $R_g$, where the gravitional radius $R_g=GM/c^2$, $i$ between 0 and $\pi/2$ and $\dot m$ between 0.001 and 0.9 times the Eddington-limit accretion rate. A distance of 94.4 Mpc was used to convert between luminosities and observed fluxes. No correction for reddening or dust extinction has been made. The calculated reprocessed light curves were resampled to match the observation epochs of the corresponding optical light curves. We subtracted the 2.54 mJy galaxy contribution (section \ref{sec:varprop}) to the $V$ band flux before fitting the reprocessed light curve to the data and subsequently added the galaxy flux to the reprocessed light curve for plotting purposes. 

For a standard thin disc \citep{shakura}, around a black hole of mass $5.24\times 10^7 M_\odot$ and with an average accretion rate of $\sim5$\% of the Eddington limit, 50\% (100\%) of the $V$ band emission is expected to arise from within 70 (600)$R_g$. As the light crossing time for one $R_g$ is $R_g/c\:\approx\:$250s, an edge-on disc would smooth the reprocessed $V$ light curve on time-scales of $\sim2\times 70 R_g/c=0.4$ days.  We note that the inclination angle to the observer is only used to calculate the correct light travel times from different locations in the disc to the observer. As light crossing time of the optically emitting regions is shorter than the usual 2-day sampling rate of the X-rays, the smoothing of the fluctuations produced by differential light crossing time is small and the reprocessed light curve does not depend sensitively on $i$. 
 
\begin{figure}
\psfig{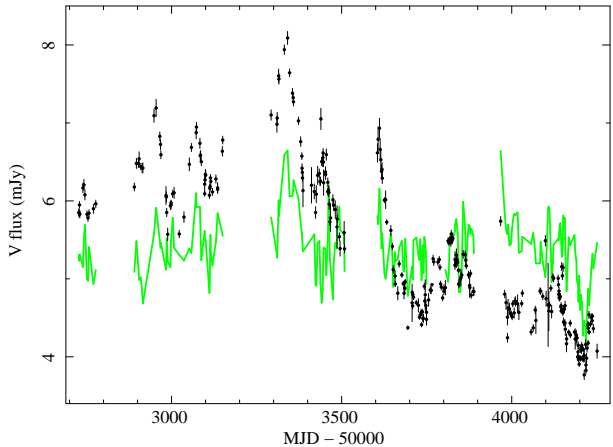}
\caption{\label{reprocess}The fluctuations in the observed $V$ band flux (black markers with error bars) cannot be reproduced by reprocessing of observed X-rays in a simple scenario where the geometry and disc accretion rate remain constant. The solid line shows an X-ray to $V$ band reprocessed light curve with disc parameters that best fit the observed $V$ band light curve.}
\end{figure}

\begin{figure}
\psfig{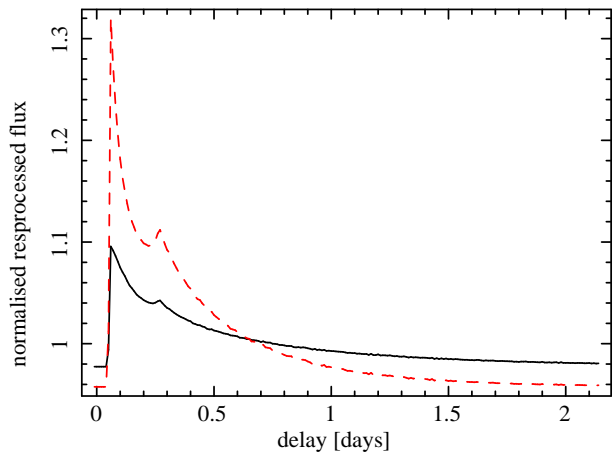}
\caption{\label{fig:transfer}Disc reprocessing transfer functions for the $V$ band (solid line) and $u$ band (dashed line) calculated using the best-fit parameters of Figure~\ref{reprocess}. The first peak corresponds to the time lag of light reprocessed at the inner edge of the thin disc on the side nearest to the observer, and the second peak to the lag of the light reprocessed at the inner disc on the far side.}
\end{figure}

Figure \ref{reprocess} shows the best-fitting reprocessed $V$ light curve together with the observed $V$ band data. The corresponding parameters are $h_X=21.5$ $i=24^\circ$, $\dot m =0.06$ and $R_{\rm in}=37.6R_g$. The mean predicted optical flux is very similar to the observed flux, the ratio of mean fluxes for the light curves shown is 0.993. The normalised transfer functions for the $V$ and $u$ bands, calculated using these best-fit parameters, are shown in Figure~\ref{fig:transfer}. It assumes an X-ray flare of observed flux 4$\times$10$^{-11}$ \ecs, illuminating the disc for 0.01 days.
The first peak in the transfer functions corresponds approximately to
the time lag of light reprocessed at the inner edge of the thin disc
on the side nearest to the observer, and the secondary peak to the light reflected at $R_{\rm in}$ on the far side. The two peaks
merge together for smaller inner truncation radii. Varying disc
parameters has a small effect on the shape of the transfer functions
and we note that in all cases their median lag is less than two days,
i.e. smaller than the X-ray sampling. Therefore, the main effect of
changing model parameters on the shape of the resulting optical light
curve is to change the relative amount of intrinsic to reprocessed
optical flux.

As can be seen from Figure~\ref{reprocess}, the reprocessed light curve reproduces the medium term fluctuations on the order of a hundred days moderately well, showing that it is energetically feasible to produce these optical fluctuations through reprocessing. The long term trend in the optical, however, is not present in the X-ray band and is not reproduced by the reprocessed light curve.  This trend in the optical suggests that either the geometry changes, affecting the amount of reprocessed X-rays or that the thin disc intrinsic luminosity changes, over time-scales of years. To test these possibilities, we fitted each segment separately, allowing only one parameter at the time ($R_{\rm in}$, $\dot m$ or $h_X$) to vary between the segments. This setup reproduces roughly the behaviour of a change in disc/corona geometry or accretion rate on time-scales of years, while all other parameters remain the same. These scenarios, where the long term trends are produced by varying one model parameter, will produce different $u$ band long-term variations. We therefore include the $u$ band light curve in the fit in an attempt to reproduce simultaneously both optical bands with the same set of model parameters over the three year segments where all three light curves overlap. 

\begin{figure}
\psfig{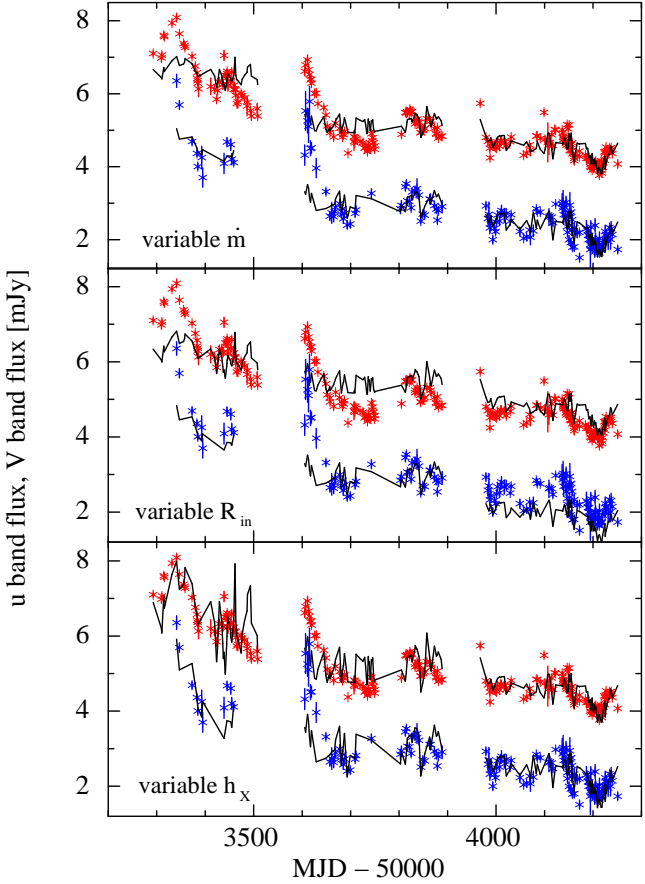}
\caption{\label{rep_tests}U (blue) and V (red) band light curves and predicted reprocessed X-ray flux (solid lines) obtained by changing only one parameter in the model between the different segments.}
\end{figure}

We start by allowing only the disc accretion rate to vary from one light curve segment to the next, while keeping the other two parameters constant between then segments. The best-fitting values of the accretion rate were 11\% of the Eddington limit for the first segment shown in Figure~\ref{rep_tests}, 4.8\% for the second segment and 0.1\% for the last segment, while keeping $R_{\rm in}=6 R_g$ and $h_X=6 R_g$. The resulting reprocessed $u$ and $V$ light curves are shown in the top panel of Figure \ref{rep_tests}. The large range in $\dot m$ is a consequence of the small truncation radius used, and represents the case where almost all of the optical emission in the lowest flux orbit comes from reprocessing. Other solutions, where $\dot m$ does not vanish, are possible, but require a larger $R_{\rm in}$ and results in greater contribution from the intrinsic disc emission. 
The middle panel on this figure shows the result of fitting the different segments with a variable $R_{\rm in}$, keeping the accretion rate and source height fixed. In this case, the accretion rate was $\dot m$=7.7\% and the source height $h_X=7.8 R_g$. The best-fitting truncation radii were $R_{\rm in}=6 R_g$ (minimum value allowed) for the first segment, $R_{\rm in}=19.6R_g$ for the second and $R_{\rm in}=41 R_g$ for the third. Finally, the bottom panel in Figure~\ref{rep_tests} shows the result of a varying source height $h_X$. The best-fitting values are $R_{\rm in}=6 R_g$, $\dot m=0.01$ and $h_X=21.3 R_g$ for the first segment, $h_X=11.8 R_g$ for the second and $h_X=6 R_g$ for the third. Evidently, varying any of these model parameters can reproduce broadly the long term trends in the $V$ and $u$ flux. The details of the reprocessed light curves, however, vary. In particular, varying the stable disc accretion rate adds different amounts of constant flux to the segments, while the amount of reprocessed flux remains approximately the same. Therefore, the short term fluctuations, arising from reprocessing, are the same size at high and low average flux levels. The opposite is true when varying geometry parameters (source height or truncation radius). In these cases the size of the short term fluctuations scale with the mean flux. The average $u$ and $V$ flux levels are best reproduced by changing the accretion rate between the segments while the worst fit is obtained by varying $R_{\rm in}$. Changing this parameter has a much stronger effect on the $u$ band than the $V$ band, so the change in $u$ flux is too large while the corresponding change in $V$ is not large enough. The best overall fit, to the shape and the average flux levels is obtained by varying $h_X$, though the quality of the fits does not allow us to strictly rule out any of the scenarios.

\begin{figure}
\psfig{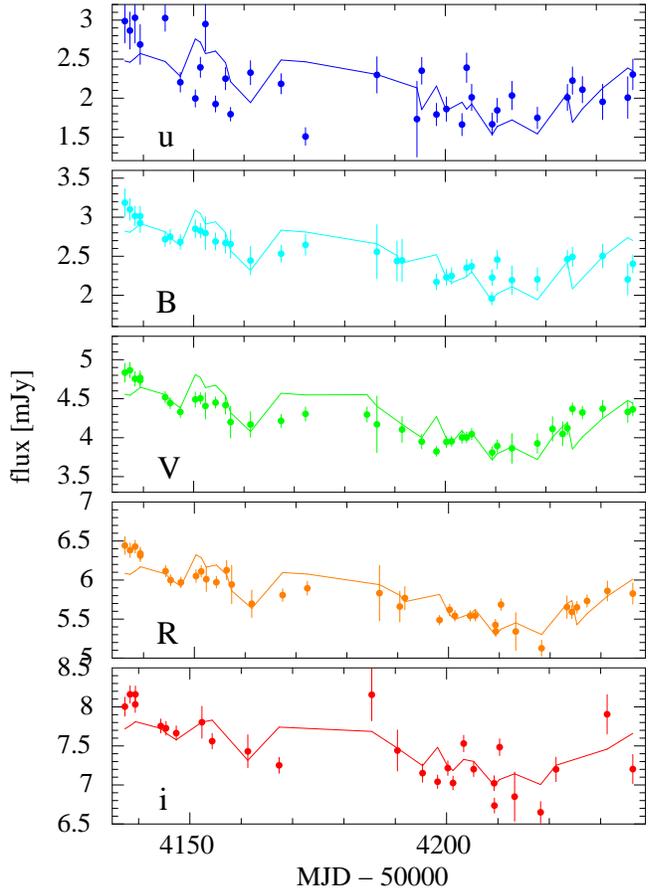}
\caption{\label{rep_intensive}Reprocessed light curves for the multicolour intensive sampling data. The model parameters were fixed to the best-fitting parameters to the $V$ band and the galaxy contribution in the other bands was allowed to vary as a free parameter. }
\end{figure}

We also fitted the multicolour light curves obtained with te Faulkes Telescope with the reprocessing model, to comfirm that this process can produce the short-term fluctuations in the different optical bands. We fixed the model parameters to the values obtained from fitting the third segment of the $u$ and $V$ light curves with either variable $h_X$ or variable $\dot{m}$, as explained above, i.e. $R_{\rm in}=6 R_g$, $\dot m=$1\% and $h_X=6R_g$. As high resolution images exist only for the $V$ band, we cannot estimate the galaxy contribution to the flux in the other optical bands, so we left the galaxy flux as a free parameter. The resulting galaxy fluxes are $u=0.27$, $B=0.67$, $V=2.54$, $R=4.08$ and $i=5.83$ mJy. The reprocessed light curves for each band are shown by the solid lines in Figure~\ref{rep_intensive}.

\section{Conclusions}
We have presented the long-term optical ($u$ and $V$ band) and X-ray
light curves of \mkn79, obtained by combining data from several monitoring programs of this source. The $u$ band light curve is the result of 2.5 years of monitoring with the Liverpool Telescope (LT) and the $V$ band light curve was constructed from the combined data from six ground based telescopes, including the LT. The X-ray observations, which were made every $\sim2$ days, are from our continuing monitoring program with the \rxte~satellite. We further included the results from a 3 month multiwavelength program
on the Faulkes Telescope in $u$, $B$, $V$, $R$ and $i$.

We summarise our findings as follows:
\begin{enumerate}
\item The light curves are variable on time-scales as short as days (X-rays) and as long as thousands of days (optical), with highly correlated variability on time-scales of tens of days or months. \item Long time-scale ($\sim$years) variations are clearly seen in the optical light curve, but are not present in the X-ray light curve.
\item Although there is more variability in the X-rays than in the optical bands on short ($\sim$day) time-scales, the fractional variability of the $V$ band light curve on long time-scales is greater when corrected for host galaxy flux, due to the large amplitude of these long-term optical variations.
\item The lags measured between the X-rays and long-term optical bands, are consistent with zero days for the data set as a whole. Formally we find that the $V$ band lags behind the X-rays by $0.00^{+1.91}_{-4.00}$ days and the $u$ band lags by $-1.95^{+1.95}_{-3.89}$ days.
\item While some strong events on a time-scale of a few days are consistent with the zero days lag, there are also time periods where the optical and the X-rays match up poorly and events are only seen in one passband.
\item The variations between the short multicolour light curves are also well correlated. The measured centroid lags, $\tau$, are consistent with the $\tau\propto\lambda^{4/3}$ prediction of X-ray reprocessing by an optically thick accretion disc. However, as the multicolour light curves are quite short, with few prominent peaks and troughs to constrain the lags, the error bars on the lag measurements are large and it is not possible to constrain the reprocessing model particularly tightly. We also cannot exclude the possibility that the measured delays across the optical bands are affected by a contribution from light reprocessed in the dusty torus further out.
\item Optical variations on time-scales of $\sim$months are well reproduced by a reprocessing model, but longer term variations ($\sim$years) cannot be accounted for in this way.
\end{enumerate}

In order to reproduce the long time-scale optical trends mentioned above, we are forced to include a second mechanism for the production of optical variations. This mechanism could be a variation in the
geometry of either the purported accretion disc or the irradiating X-ray source. Changing the height of the X-ray source changes the fraction of the disc illuminated and hence the amount of optical emission produced by reprocessing. The intrinsic disc flux produced by viscous dissipation, is governed by $\dot m$ and changes in the inner truncation radius $R_{\rm in}$ affects the amount of optical emission produced. We have considered the effect of changing $h_X$, $\dot m$ and $R_{\rm in}$ individually, in steps between the three sections of the optical light curves (Figure \ref{rep_tests}). We find that when we vary $h_X$, we are able to reproduce the large and small fluctuations of the observed light curves slightly better than when we vary the disc parameters. It is not possible, however, at this level of modelling, to rule out the other two possibilities. A crude model like this cannot reproduce the observed light curves exactly. In reality the parameters will probably all vary together, rather than one at a time as we have done here, and also gradually, rather than in a step-wise fashion. The description of the X-ray source as a point source above the centre of the disc, although providing a reasonable approximation to an extended hemispherical corona, may also not be correct. However,
the requirement for an additional source of optical variation, in addition to X-ray reprocessing, is robust.

Despite the good correlation on short time-scales, there are some short time-scale variations which appear to occur only in one passband. This mismatch needs to be addressed by future studies and may be a consequence of anisotropic X-rays emission \citep[see, e.g.][]{gaskell06}.

Within the standard accretion disc model \citep{shakura}, the high black hole mass and low accretion rate of \mkn79 implies that it has a cool disc. In such a disc, the X-ray and optical emitting regions are closer together, in terms of gravitational radii, than in a system of low black hole mass and/or high accretion rate.
Hence both reprocessing and intrinsic variations are expected to contribute to the observed optical variability. Reprocessing benefits from the large solid angle subtended by the disc in the inner regions and intrinsic variations propagate on observable time-scales, because of the shorter viscous time-scale at smaller disc radii. We have previously reported this behaviour in the quasar MR2251-178 \citep{arevalo08}. This system has a very massive central black hole ($M\sim10^9M_\odot$) and hence a cool disc, with the X-ray and optical emitting
regions even closer together than in \mkn79, in terms of gravitational radii. When  accretion rate fluctuations propagate through the disc, the X-ray and optical emitting regions are modulated together, leading to a good correlation on long time-scales. Reprocessing of X-rays imprints small rapid fluctuations onto the optical light curve, resulting in a well correlated behaviour on short time-scales as well. 
The strong X-ray--optical correlation we have reported here is in agreement with this picture.

For \mkn79 we expect that 50\% of the $V$ band emission
should originate within 70$R_g$ of the black hole. Taking a typical
value for the viscosity parameter, $\alpha\sim 0.1$, in an optically
thick, standard accretion disc \citep{shakura}, with thickness $(h/R)^2 \sim 0.01$, 
we find that the viscous time-scale at 70$R_g$ is $t_{\mbox{\small{visc}}}\sim 5$ years. In a geometrically thin disc, this time-scale is even longer.  This (approximate) time-scale,
on which we would expect accretion rate perturbations to propagate
inwards to the X-ray emitting region, is longer than the time over which we have observed the steep decrease in optical flux and hence  consistent with the fact that we have not yet seen a decrease in
long term average X-ray luminosity. However the connection between the
disc and the corona is not well understood and it is not clear how
accretion rate variations, which may affect the optical disc emission,
will be translated to the X-ray emission. If this interpretation is correct,
the X-ray light curve should decline significantly over the next 5 or more years.

In subsequent papers we will compare the optical and X-ray variations
in a number of other AGN of differing black hole mass and accretion
rate, and hence of differing disc structure, in order to determine how
disc structure affects both the long and short term optical
variability of AGN.

\section*{Acknowledgements}
We thank the staff of the Rossi X-ray Timing Explorer for their usual
extremely helpful and highly efficient scheduling of our observations.

We also thank the Las Cumbres Observatory Global Telescope for
observations obtained with the Faulkes Telescope North, Hawaii, and the
staff of the Liverpool Telescope for their continued support of our
programme. The Liverpool Telescope is operated by Liverpool John Moores
University with financial support from the UK Science and Technology
Facilities Council.

EB gratefully acknowledges financial support in the form of a Stobie-SALT
Scholarship from South African National Research Foundation and the
University of Southampton. PA and IMcH acknowledge support from the STFC
under rolling grant PP/D001013/1 and PU acknowledges support from an STFC
Advanced Fellowship. EMC gratefully acknowledges support provided by NASA
through the Chandra Fellowship Program, grant number PF8-90052.

We are grateful to Constantin Grankin for making observations with the
0.6-m telescope at Mt. Maidanak, and to Evelina Gaynulina for the data
reduction. We also wish to thank Jeff Campbell, Taylor Chonis, Stephanie 
Gilbert, Cece Hedrick, Liz Klimek, Amanda Kruse, Shoji Masatoshi, Tim Miller,
Eric Petersen, Aaron Watkins and the late David Brokofsky for making
observations with the University of Nebraska 0.4-m telescope, and also A.
J. Benker, James Wallace, and Kate Wheeler for their help with reduction
of the observations. Observations at the University of Nebraska and at Mt.
Maidanak Observatory have been supported by the US National Science
Foundation through grants AST 03-07912 and AST 08-03883, and by the
University of Nebraska Layman Fund and Pepsi-Cola UCARE program.

\bibliographystyle{mn2e}
\bibliography{../library} 
\label{lastpage}

\end{document}